\documentclass[pra,twocolumn,showpacs,superscriptaddress]{revtex4}

\usepackage{graphicx}

\newcommand{\nn}{\nonumber}

\begin{document}

\title{Axisymmetric Vortices in Spinor Bose-Einstein Condensates under
Rotation}
\author{Tomoya Isoshima}
\email{tomoya@focus.hut.fi}
\affiliation{
Materials Physics Laboratory,
Helsinki University of Technology, \\
P.~O.~Box 2200 (Technical Physics), FIN-02015 HUT, Finland
}
\author{Kazushige Machida}
\affiliation{
Department of Physics, Okayama University, Okayama 700-8530, Japan
}

\date{\today}
\begin{abstract}
The relative stability of various axisymmetric vortices in a
spinor Bose-Einstein condensate with $F=1$ is
examined within extended Bogoliubov theory. This yields the phase
diagram in the plane of external rotation frequency vs magnetization.
We compare antiferromagnetic, nonmagnetic, and ferromagnetic cases.
The excitation spectrum is evaluated under rotation
to investigate the local stability of the possible vortices
and the vortex nucleation frequency.
\end{abstract}

\pacs{03.75.Fi, 67.57.Fg}

\keywords{
Vortex, spinor BEC,
Three component Bose-Einstein condensation,
Gross-Pitaevskii equation, Bogoliubov equation,
Collective modes
}

\maketitle

\section{Introduction}
Much attention has been focused on Bose-Einstein condensation (BEC)
realized in atomic gases~\cite{firstRb,hulet,ketterle,pethick}.
The atomic species investigated include the isotopes
$^{87}$Rb and $^{85}$Rb and also $^{23}$Na, $^{7}$Li, H, $^{4}$He, and $^{41}$K
(see Refs.~\cite{firstRb,cornish,ketterle,pethick,hulet,fried,robert,modugno}).
%
%
%
%
%
These systems provide one with unique opportunities to investigate novel
states of superfluid matter waves.
When an atomic gas is cooled in an applied external magnetic field,
the condensate is described by a scalar order parameter.
The magneto-optic trapping
method  is widely used to study BEC.
Recently it has become possible
to trap an atomic gas using purely optical methods,
thus the resulting BEC retains its original atomic hyperfine state.
Specifically, $^{23}$Na~\cite{stenger}
and  $^{87}$Rb~\cite{barrett}
both with $F=1$ are successfully Bose-condensed via optical trapping.
This system, dubbed spinor BEC, is now described by multicomponent
order parameter.

Ohmi and Machida~\cite{ohmi}, and Ho~\cite{ho} have independently
introduced the basic Hamiltonian for describing this spinor BEC by
extending the Bogoliubov 
formalism to study the fundamental
properties of this interesting multicomponent BEC, pointing out the
richness of
the topological defect configurations such as the $l$-vector textures and
domain-wall structures.
These are analogous to those found in the A and B phases of
superfluid $^3$He~\cite{salomaa_vtx,salomaa_wall_b,salomaa_wall_a}.
An advantage of the dilute BEC systems with respect to strongly
interacting liquid helium (for $^4$He, there exists no microscopic theory;
for $^3$He, there is a BCS-like microscopic theory,
generalized for {\it p}-wave pairing) is that for the dilute gas one can 
make controlled approximations, treating the interparticle
interactions as small pertubative parameter.
Moreover, it is possible to directly visualize the condensate in atomic gases
using optical methods.
Spin of condensate is easily controlled by external magnetic field.
The BEC systems are quite versatile also in that the interaction
parameter can be adjusted over a large range --- even its sign can be changed.
Finally, the condensate in BEC systems can exhibit several spin states,
{\it e.g.}, with the hyperfine spins $F=1$ and $F=2$.

Theoretical studies of vortices and other topological defects
in spinor BECs was 
initiated by Ohmi and Machida~\cite{ohmi} and Ho~\cite{ho}.
Systematic investigations on vortices were 
followed by Yip~\cite{yip} who considered both axisymmetric and
non-axisymmetric vortices,
and by Isoshima {\it et al.}~\cite{isoshima}, who only considered
axisymmetric vortices and their excitation spectra. 
Leonhardt and Volovik~\cite{leonhardt}, Stoof~\cite{stoof},
Merzlin {\it et al.}~\cite{marzlin}, Zhou~\cite{zhou} and Martikainen and
Suominen~\cite{martikainen}
examined exotic topological defect
structures in spinor BECs.

Here we continue our studies of spinor BECs~\cite{ohmi,
spindomain,isoshimanakahara1,isoshimanakahara2,double}; in
particular, those on vortices with $F=1$~\cite{isoshima}.
We investigate the vortex phase diagram
in the plane spanned by the magnetization (which
is given {\it a priori} when a three-component atomic gas is
prepared) and the external rotation.
We consider both the antiferromagnetic and ferromagnetic cases.
The former is realized in $^{23}$Na while the latter is expected
for $^{87}$Rb.
The nonmagnetic situation in which the spin channel interaction $g_s=0$ is
studied
as the limiting case for $|g_s|\ll g_n$
($g_s$ and $g_n$ are the interaction constants for the spin and density channels).
We restrict our calculations to axisymmetric vortices with winding
numbers less than or equal to unity~\cite{isoshima}.

The organization of this paper is as follows:
After giving a brief introduction to the Hamiltonian for the system and
the extended Gross-Pitaevskii equation,
we enumerate the possible vortices
allowed by axisymmetry in Sec. 2.
To investigate the global stability of the various vortex structures,
the relative free energies of the different vortices are
compared as functions of magnetization and rotation frequency in
Sec. 3.
Section 4 presents the excitation spectra for each vortex by
solving the associated Bogoliubov equations extended
to an order parameter with three components
in order to investigate whether a given vortex state
is stable against collective modes.
This yields a local stability criterion for each vortex type.
This consideration extends our previous works of nucleation
criteria~\cite{isoshima1,isoshima2,isoshima_T,isoshima4,mizushima}.
The final Section presents a summary and discussion.


\section{Possible Types of Axisymmetric Vortices}

We treat the system of a Bose condensate with internal
degrees of freedom $F=1$.
Hence the condensate order parameter is characterized
by three components with $m_F=1,0,-1$.
External rotation of the system around the rotation axis,
perpendicular to the disc-shaped two-dimensional plane, is denoted by the
angular velocity
$\Omega$ which has a sense ($+$ or $-$).

We start with the system Hamiltonian~\cite{ohmi,ho}
\begin{eqnarray}
   H &=&
   \int\! d{\bf r} \biggl[
      \sum_{j}\Psi_j^{\dagger}
          \left\{ - C \nabla^2 + V({\bf r}) - \mu_j \right\}
      \Psi_j
\nn\\&&\  %
      + \frac{g_n}{2} \sum_{jk}
         \Psi_j^{\dagger} \Psi_k^{\dagger} \Psi_k \Psi_j
\nn\\&&\  %
      + \frac{g_s}{2} \sum_{\alpha}
          \sum_{jklm} \Psi_j^{\dagger}\Psi_k^{\dagger}
          (F_{\alpha})_{jl}(F_{\alpha})_{km}
          \Psi_l \Psi_m
\nn\\&&\  %
      - {\bf \Omega} \cdot \sum_{j} \Psi_j^{\dagger}({\bf r} \times {\bf
p})\Psi_j
   \biggr]
\label{eq:ham}
\end{eqnarray}
with $C = \hbar^2/(2m_a)$. The interaction is  characterized by the
two kinds of channels; the density channel:
$g_n = 4 \pi \hbar^2 (a_0 + 2a_2)/(3 m_a)$, and
the spin channel:
$g_s = 4 \pi \hbar^2 (a_2 - a_0)/(3 m_a)$.
The atomic mass is $m_a$.
The scattering lengths $a_0$ and $a_2$ characterize
collisions between atoms with total spin 0 and 2.
The subscripts are $\alpha = (x,y,z)$ and $i,j,k,l = (0, \pm1)$.
The latter correspond to the above three species.
The scalar field $V({\bf r})$ is the harmonic confining potential.
The angular-momentum matrices $F_\alpha$ are
\begin{eqnarray}
 F_x &=& \frac{1}{\sqrt{2}}\left(
   \begin{array}{ccc}
       0 & 1 & 0 \\ 1 & 0 & 1 \\ 0 & 1 & 0
   \end{array}\right),
\nn \\
 F_y &=& \frac{i}{\sqrt{2}}\left(
   \begin{array}{ccc}
       0 & -1 & 0 \\ 1 & 0 & -1 \\ 0 & 1 & 0
   \end{array}\right),
\\
 F_z &=& \left(
   \begin{array}{ccc}
       1 & 0 & 0 \\ 0 & 0 & 0 \\ 0 & 0 & 1
   \end{array}\right).
\nn
\end{eqnarray}

The chemical potentials $\mu_i$ obey $\mu_1 - \mu_0 = \mu_0 - \mu_{-1}$.
We introduce $\mu = \mu_0$ and $\mu^{\prime} = \mu_1 - \mu_0$.
The Gross-Pitaevskii equation for this system, extended to the three
components becomes
\begin{eqnarray}
    \sum_j\biggl[
        \left\{
            -C \nabla^2 + V(r) - (\mu + \mu^{\prime}j)
        \right\} \delta_{jk}
&&\nn\\
        + g_n \sum_l |\phi_l|^2 \delta_{jk}
&&\nn\\
        + g_s \sum_{\alpha}(F_\alpha)_{jk} \sum_{lm} (F_\alpha)_{lm}
            \phi_l^{\ast} \phi_m
&&\nn\\
        - i \hbar \Omega \cdot \nabla \times {\bf r} \delta_{jk}
    \biggr] \phi_k
    &=&  0.
\label{eq:gp}
\end{eqnarray}
The total energy of the condensate is given by
\begin{eqnarray}
E &=& \int d^2 {\bf r}
   \biggl[ \sum_j \phi_j^{\ast} \left(
       -C \nabla^2 + V(r)
   \right) \phi_j
\nn\\&&
\     + \frac{g_n}{2} \sum_{jk} |\phi_j|^2 |\phi_k|^2 + E_s
\nn\\&&
\     - i \hbar {\bf \Omega} \cdot \sum_{j}  \phi_j^{\ast} \nabla \phi_j
\times {\bf r}
   \biggr],
\label{eq:E}
\end{eqnarray}
where
\begin{eqnarray}
 E_s &=&
      \frac{g_s}{2} \sum_\alpha \left(
           \sum_{jk} \phi_j^\ast (F_\alpha)_{jk} \phi_k
      \right)^2.
\label{eq:Es}
\end{eqnarray}

\begin{figure}[htbp]
\begin{center}\leavevmode
\includegraphics[width=7cm]{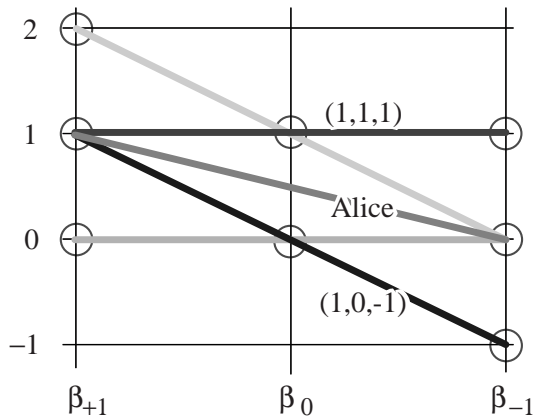}

\caption{\label{fig:windings}
Possible allowed vortex configurations under axisymmetry.
Each condensate $\phi_{i}$ ($i=1,0,-1$) has the phase $\beta_{i}$.
The vortices; (1,0,-1), (1,$x$,0) and (1,1,1) are treated in this paper.
}
\end{center}
\end{figure}

Since we treat a cylindrically symmetric disc-shaped two-dimensional system,
the condensate wavefunction $\phi_j$ may be decomposed in the form:
\begin{eqnarray}
   \phi_j(r, \theta) &=& \phi_j(r) \gamma_j(\theta)
\nn\\
 &=& \phi_j(r) \exp[ i (\alpha_j + \beta_j \theta) ]
\end{eqnarray}
where the condensate wavefunctions $\phi_j({\bf r})$ ($j = 0, \pm 1$) are
expressed in terms of cylindrical coordinates.
The phases of the condensate wavefunctions are determined
such that the energy $E_s$ in Eq.\ (\ref{eq:Es}) is minimized.
The minimizing of the above spin-dependent term:
\begin{eqnarray}
    E_s(r)
&=&
    \frac{g_s}{2} \biggl[
        2 \phi_{0}^2(r) \{
            \phi_{1}^2(r) + \phi_{-1}^2(r) 
\nn\\&&
\         + \phi_1(r) \phi_{-1}(r) \left(
                \gamma_1 \gamma_{-1} \gamma_0^{\ast 2} +
                \gamma_{1}^\ast  \gamma_{-1}^\ast \gamma_0^{2}
            \right)
        \}
\nn\\&& 
\       +  \left(  \phi_{-1}^2(r) - \phi_{1}^2(r)  \right)^2
    \biggr]
    \label{eq:gsterm}
\end{eqnarray}
leads to the condition 
\begin{equation}
    \gamma_1 \gamma_{-1} \gamma_0^{\ast 2} = \pm1,
    \label{eq:phase}
\end{equation}
where the upper (lower) sign is used for the ferromagnetic (antiferromagnetic) case 
$g_s < 0$ ($g_s > 0$), respectively.
This condition may be rewritten in terms of $\alpha$ and $\beta$ as
\begin{eqnarray}
    2 \alpha_0 &=& \alpha_1 + \alpha_{-1} + n \pi
\\
    2 \beta_0 &=& \beta_1 + \beta_{-1}
\end{eqnarray}
where $n$ is an integer.
We take $\alpha_{\pm 1} = \alpha_0 = 0$ in the following
since they have no effect in the discussion below.
The phases of the three components are now 
expressed in the form
\begin{equation}
   \left(
      \begin{array}{c} \gamma_1 \\ \gamma_0 \\ \gamma_{-1} \end{array}
   \right) =
   \left(
      \begin{array}{c}
         \exp(i \beta\theta) \\
         1 \\
         \pm \exp(- i \beta \theta) \\
      \end{array}
   \right)
   \exp(i \beta_0 \theta)
\label{eq:gammas}
\end{equation}
using $\beta \equiv \beta_1 - \beta_0 = \beta_0 - \beta_{-1}$.
In Fig. \ref{fig:windings}, the various lines show the possible combinations
of the $\beta_i$,
giving rise to the vortex types which we consider in the  following.
If we restrict our consideration to the phase coefficient, or the winding
number being less than 2,
there are nine possible combinations of the $\beta_i$.
There is one vortexfree configuration and
the other combinations are  (1,0,-1), (1,1,1), and $(1,\frac{1}{2},0)$ where the
winding numbers
of the three components $\phi_1$, $\phi_0$, and $\phi_{-1}$ are denoted in
this order.
The latter vortex configuration is only realized in a form (1,$x$,0) where
$\phi_0$ vanishes (where $\beta_i$ is fractional).
It should be noted that the restriction for the winding number to be less than 2
is reasonable,
because a vortex with a higher winding number is unstable
and easily breaks up into vortices with unit winding number.



%
%
%
%

\section{Relative Stability of Various Vortices}

We compare the relative energies of the various vortices enlisted above.
There are eight candidates for the possible vortex type, allowed by
axisymmetry.
Three of them are (1,0,-1), (1,1,1), and (1,$x$, 0) where ``$x$'' denotes that this
component vanishes identically as mentioned above.
The last one is named Alice vortex after the Alice string~\cite{leonhardt}.
By interchanging the roles of $\phi_{1}$ and
$\phi_{-1}$, we can obtain two more vortices: (-1,0,1), and (0,$x$,1).
The remaining combinations (0,$x$,1),(-1,$x$,0) and (-1,-1,-1) always have
higher energies than
those of (0,$x$,1), (1,$x$,0) and (1,1,1), respectively, when the external angular
velocity $\Omega$ is positive.
Thus, in what follows, we consider the vortices, (1,0,-1), (1,1,1), (1,$x$,0), (-1,0,1), and (0,$x$,1) for
$\Omega\geq 0$ and $M/N\geq 0$.

The mass of a $^{87}$Rb atom is $m_a = 1.443 \times 10^{-25} {\rm kg}$.
This value is used in the following.
We use the scattering lengths
$a_0 = 5.5\times 10^{-9}\,{\rm m}$ and $a_2 = 5.843 \times 10^{-9}\,{\rm m}$
for the antiferromagnetic case.
The ratio of the interaction coefficients is $g_s / g_n = +0.02$.
Another set for the ferromagnetic case is
$a_0 = 5.5\times 10^{-9}\,{\rm m}$ and
$a_2 = 5.182 \times 10^{-9}\,{\rm m}$, whose
ratio of the interaction coefficients is $g_s / g_n = -0.02$.
The coefficient of the magnetic term vanishes when
$a_0 = a_2 = 5.5\times 10^{-9}\,{\rm m}$.  The ratio becomes $g_s / g_n = 0$.
The amplitude of the ratio $|g_s / g_n| = 0.02$ is for a Na atom.

For each vortex type and the $g_s$ and $\mu^\prime$, 
the chemical potential $\mu$ is
determined such that the linear density of the particle number
becomes $2 \times 10^3 (\mu {\rm m})^{-1}$.
We use a harmonic potential with $\nu = 200 \, {\rm Hz}$
for radial confinement.
All the energies are scaled by the trap frequency $h\nu$.
The angular velocity $\Omega$ is normalized by $2 \pi \nu $ below.

The spatial profile of the condensate wavefunctions in each vortex is
determined by the above GP Eq.\ (\ref{eq:gp}), which is expressed as
\begin{eqnarray}
    \sum_k \biggl[
        \{ -C\nabla^2 + V(r)
        - \left( \hbar \Omega \beta_0 + \mu \right)
&&\nn\\
        - j \left( \hbar \Omega \beta + \mu^{\prime} \right)
        + g_n  \sum_{l} |\phi_l|^2
n \}\delta_{jk}
&&\nn\\
        + g_s \sum_\alpha  (F_\alpha)_{jk}\sum_{lm}
(F_\alpha)_{lm}\phi^{\ast}_l \phi_m
    \biggr]\phi_k &=& 0.
\label{eq:gp_r}
\end{eqnarray}
When we vary $\Omega$ and $\mu^\prime$,
the condensate wavefunctions $\phi_i$ are
determined by a single parameter $\hbar\Omega\beta + \mu^{\prime}$
because the variation of $\hbar \Omega \beta_0$ is
canceled by that of the chemical potential $\mu$.

The energy of the system Eq.\ (\ref{eq:E}) is rewritten as
\begin{eqnarray}
E&=&
    \int d^2 {\bf r} \biggl[
        \sum_j \phi_j^{\ast} \left(
            -C \nabla^2 + V(r)
        \right) \phi_j
\nn\\&&
        + \frac{g_n}{2} \sum_{jk} |\phi_j|^2 |\phi_k|^2 + E_s(r)  \biggr]
    -  \sum_j \hbar \Omega(\beta_0 + j \beta) N_j
\nn\\&=&
   E_{\rm inner} - \hbar \Omega L,
\label{eq:E:disc}
\end{eqnarray}
where the number of particles in the $j$'th component is
 $N_j = \int d^2 {\bf r} |\phi_j|^2$.
The total particle number in the system is
 $N = \sum_j N_j$,
and the total angular momentum is
 $L = \sum_j (\beta_0 + j \beta)N_j$.
The energy $E_{\rm inner}$, the magnetization
$M = \sum_j j N_j$, and $L$ are functions of the 
$\phi_j$ and, therefore, are functions of $\hbar\Omega\beta + \mu^{\prime}$.
The critical angular velocity $\Omega_c$ for a given magnetization between
two vortex types (1) and (2) is simply determined by
\begin{eqnarray}
   \hbar\Omega_c (M)
   &=&
   \frac{ E_{\rm inner}^{(1)} - E_{\rm inner}^{(2)} }{ L^{(1)} - L^{(2)} }.
\label{global_omega_c}
\end{eqnarray}
The range of $M$ which a vortex state can take
depends on the interaction constant $g_s$ and the winding number.
For example, the (1,1,1) vortex does not have an intermediate value of $M/N$
in the cases with $g_s = 0$ (nonmagnetic case) and $- 0.02 g_n$
(ferromagnetic case).
This behavior is well known for the uniform spinor condensate~\cite{ohmi,ho}.
When $g_s = - 0.02 g_n$,
the magnetization range $-1 < M/N < 0$ is not allowed for
the (1,$x$,0) vortex.
This range may be improved with 
improved numerical accuracy~\cite{virtanen,isoshima_T}.

\begin{figure}[htbp]
\begin{center}\leavevmode
\includegraphics[width=7cm]{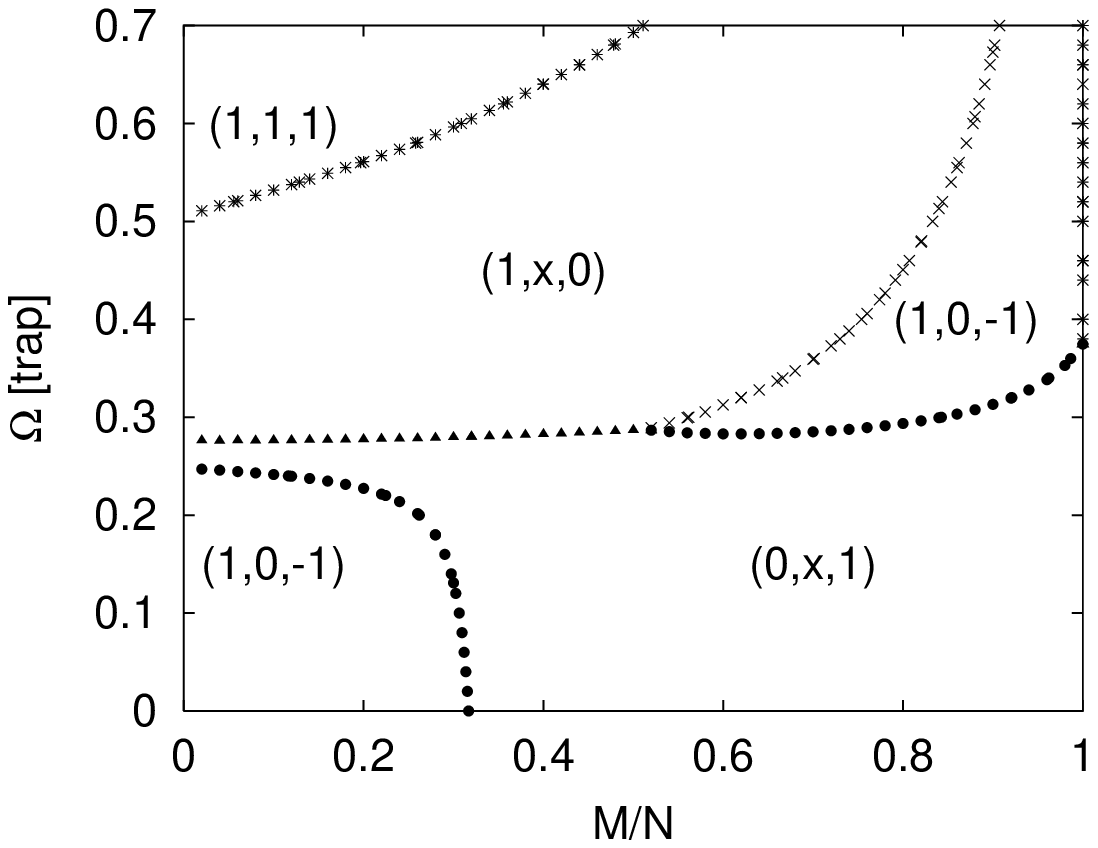}\\
(a)

\includegraphics[width=7cm]{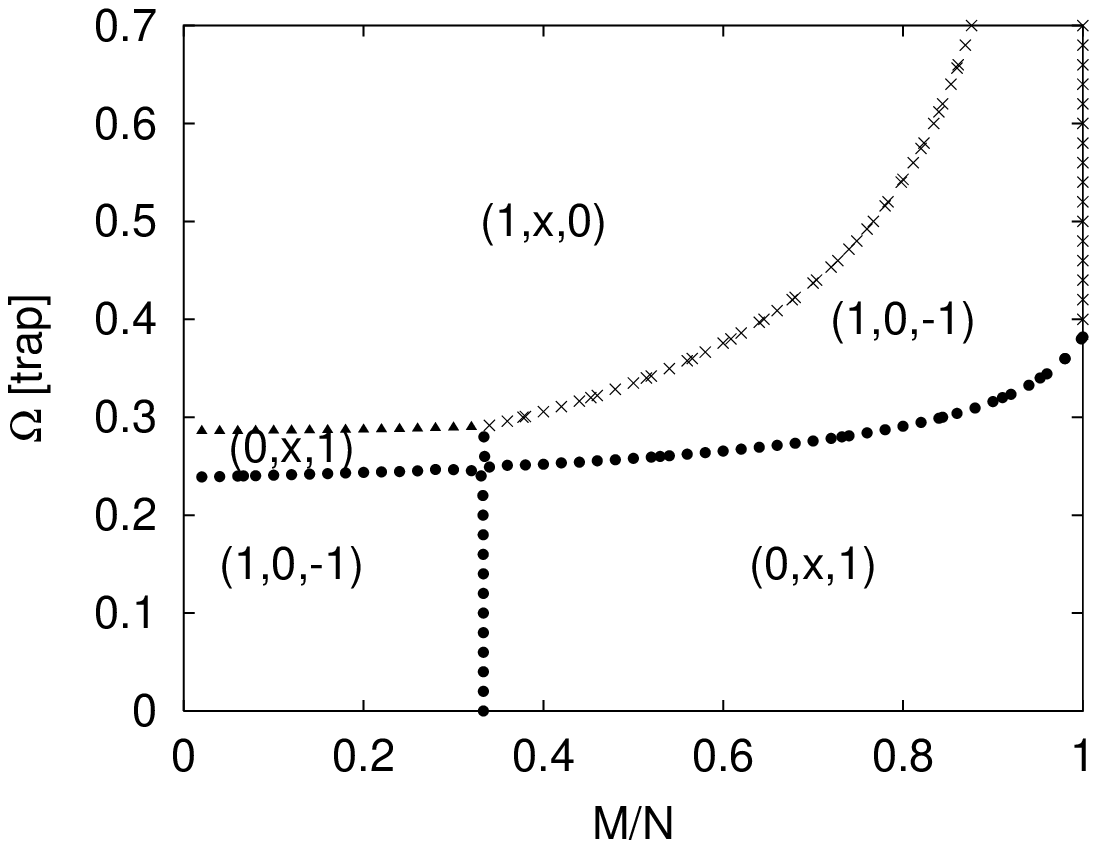}\\
(b)

\includegraphics[width=7cm]{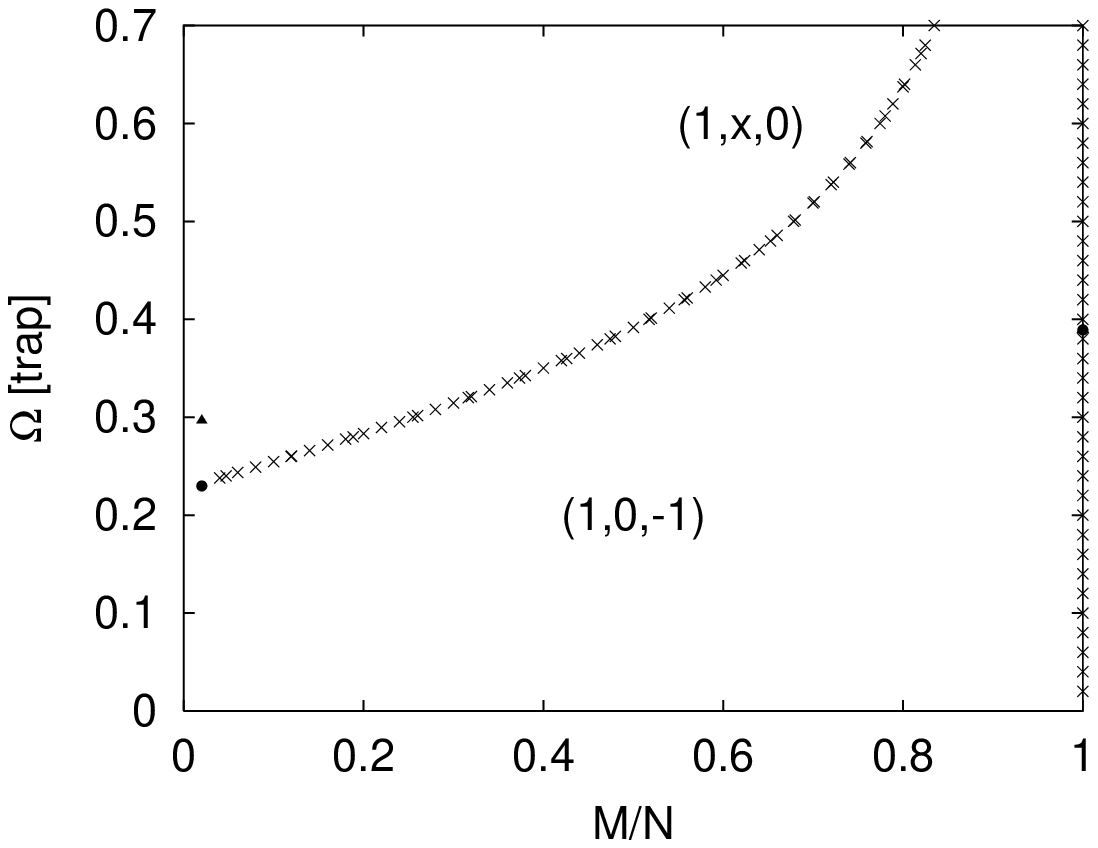}\\
(c)
\end{center}

\caption{
   \label{fig:global}
Phase diagrams in the $\Omega$ vs $M/N$ plane, showing the vortex type
which has the lowest energy $E(\Omega)$.
(a) Antiferromagnetic case ($g_s = 0.02g_n$).
(b) Nonmagnetic case ($g_s = 0$).
(c) Ferromagnetic case ($g_s = -0.02 g_n$).
 The (1,1,1) vortex is absent from the phase diagrams (b) and (c)
 because this vortex cannot have a stable configuration for $-1<M/N<1$.
 The (0,$x$,1) Alice vortex is not stabilized and
it is thus excluded from the phase
diagram (c).
}
\end{figure}

Figure \ref{fig:global} shows the phase diagrams with the  lowest energy
in the plane of $M/N$ and $\Omega$ for the three cases; the
antiferromagnetic ($g_s= 0.02g_n$),
the nonmagnetic ($g_s=0$), and the  ferromagnetic cases ($g_s = -0.02g_n$).
The critical angular velocity $\Omega_c(M)$ in Eq.\ (\ref{global_omega_c})
determines the boundaries of each domain.
When the system is antiferromagnetic or nonmagnetic,
the (0,$x$,1) vortex has the lowest energy around $\Omega=0$ for $M/N=1$.
This is because the (0,$x$,1)
vortex almost reduces to the vortexfree single-component state without the
vortex winding
in the full polarization limit.

The (1,0,-1) vortex is lowest in energy around $\Omega = 0$ with
$M/N=0$, irrespective of the magnetism.
The condensate simply becomes a vortexfree system for the ferromagnetic case.
For the antiferromagnetic case,
the nonrotating $\phi_0$ component without winding number dominates and
the $\phi_1$ and $\phi_{-1}$ components which have opposite winding numbers
are small.

This (1,0,-1) vortex and the Alice vortex compete
in the region around $\Omega \sim 0.6, M/N \sim 0.8$.
The curve between them reflects the kinetic energy of the (1,0,-1) vortex
when there is no magnetic term ($g_s=0$).
The amplitudes of the condensate wavefunctions $\phi_1(r)$ and $\phi_0(r)$
of the (1,0,-1) vortex
for the magnetization $M/N(>0)$
are equal to  $\phi_1(r)$ and $\phi_{-1}(r)$ of the (1,$x$,0) vortex for
the magnetization $2M/N-1$.
The (1,0,-1) and the (1,$x$,0) vortex
have the angular momenta $L = M$ and $L = (M + N)/2$, respectively.
The critical $\Omega_c$ in Eq.\ (\ref{global_omega_c}) becomes
\begin{equation}
  \hbar \Omega_c(M) =
   2 \frac{E_{\rm inner}^{\prime}(M) - E_{\rm inner}^{\prime}((M + N)/2)}
   {M - N}
\end{equation}
where $E_{\rm inner}^{\prime}(M)$ is $E_{\rm inner}$ of the (1,0,-1) vortex
for $0<M<N$.
The behavior of $\Omega_c$ is still complicated because
both the numerator and the denominator approach zero when $M$ approaches $N$.
%
%
%
%


\section{Local Stability}

There is another stability criterion for a vortex.
The excitation spectrum for a stable vortex must be positive definite.
In other words, if the lowest excited state in the spectrum
becomes negative as a function of $\Omega$,
the given vortex becomes locally unstable.

\subsection{Excitation spectrum}

The excitation spectrum is obtained by solving the following Bogoliubov
equation extended to the three-component BEC case:

\begin{figure}[htbp]
  \begin{center}\leavevmode

     \includegraphics[width=7cm]{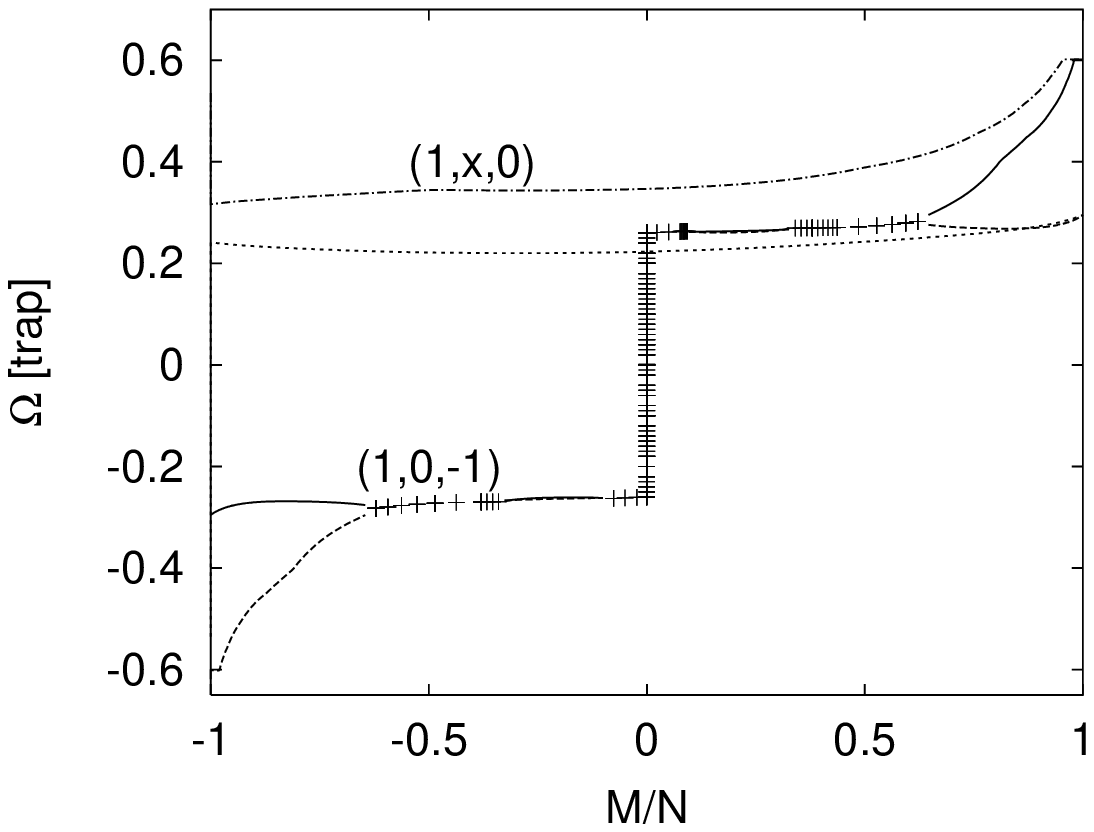}\\
     (a)\\

      \includegraphics[width=7cm]{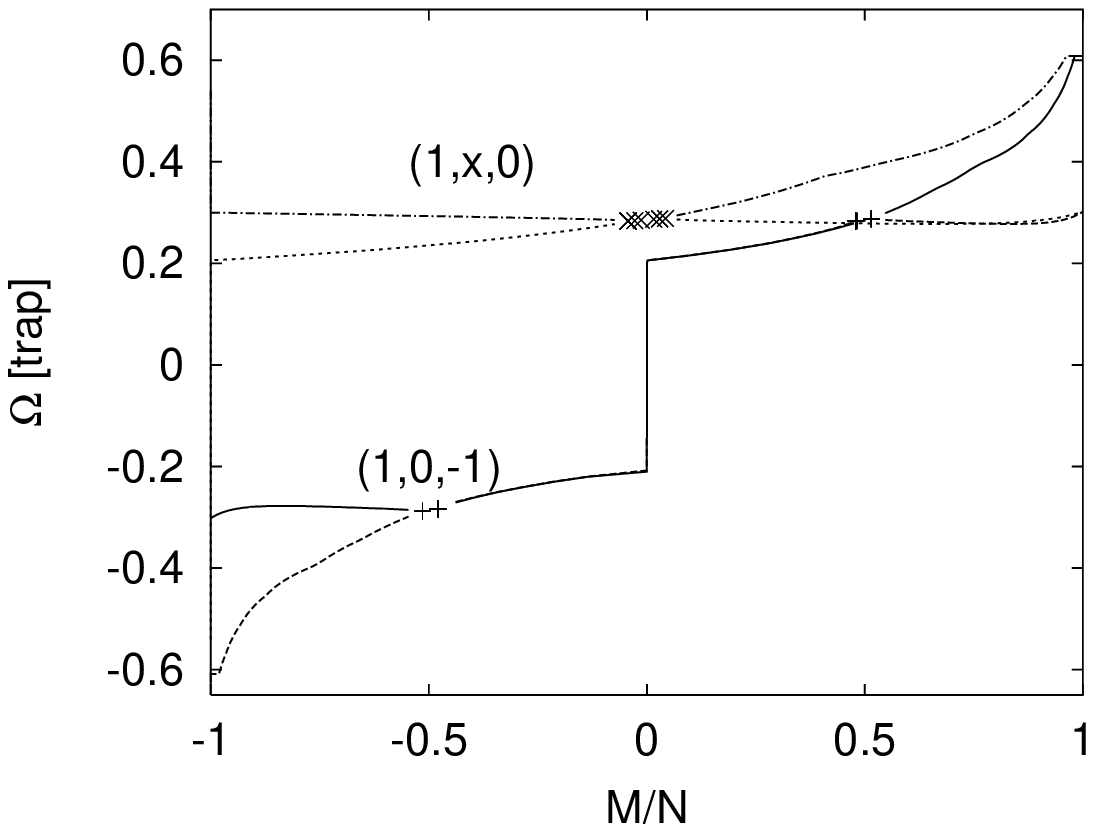}\\
     (b)\\

     \includegraphics[width=7cm]{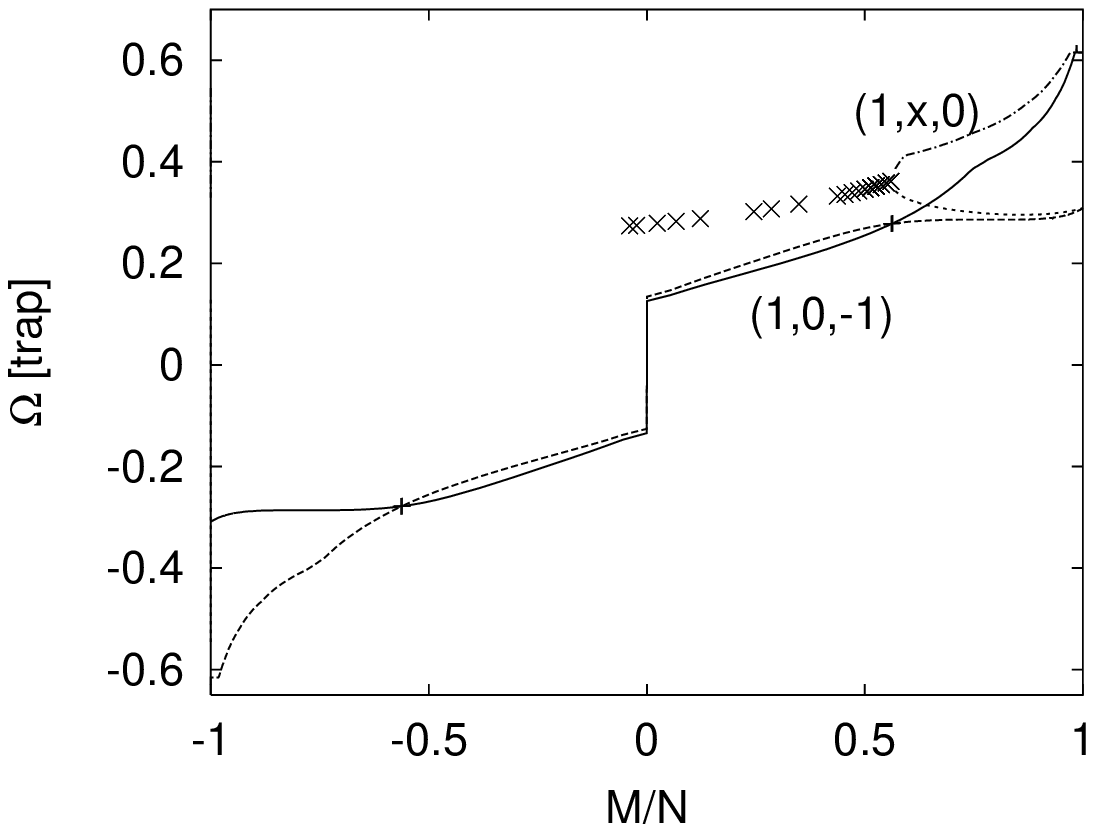}\\
     (c)
 \end{center}

\caption{\label{fig:localomg}
Local stability regions bounded by $\Omega_c^{\pm}$
for the two types of vortices, (1,0,-1) and (1,$x$,0).
(a) Antiferromagnetic case ($g_s = 0.02g_n$).
(b) Nonmagnetic case ($g_s = 0$).
(c) Ferromagnetic case ($g_s = -0.02 g_n$).
The solid line and the dashed line show
the upper critical $\Omega_c^{+}$ and
the lower critical $\Omega_c^{-}$ for a $(1,0,-1)$ vortex.
The dashed-dotted line and the dotted line show
the upper critical $\Omega_c^{+}$ and 
the lower critical $\Omega_c^{-}$ for the $(1,x,0)$ Alice vortex.
The lower critical $\Omega$ is higher than the upper critical $\Omega$
for $-0.5 < M/N < 0.5$ in the (1,0,-1) vortex.
The range  $-1 < M/N < -0.1$ for the (1,$x$,0) vortex is inhibited
because of the ferromagnetic interaction.
When the imaginary part of $\varepsilon$ is finite,
its real part is used in Eqs.\ (\ref{omega_plus}) and
(\ref{omega_minus}).
These regions are indicated with the crosses and pluses.
}
\end{figure}

%
%
\begin{eqnarray}
   \sum_k \{
        A_{jk}     u_q({\bf r}, k)
      - B_{jk}     v_q({\bf r}, k)\} &=& \varepsilon_q u_q({\bf r}, j),
\nn\\
   \sum_k \{
        B^{\ast}_{jk}   u_q({\bf r}, k)
      - A^{\ast}_{jk}   v_q({\bf r}, k)\} &=& \varepsilon_q v_q({\bf r}, j)
\label{eq:bogo1}
\end{eqnarray}
where
\begin{eqnarray}
   A_{jk} &=&
      \{-C \nabla^2 + V(r) - ( \mu + j \mu^{\prime} )
      \}\delta_{jk}
\nn\\&&
      + g_n \left\{
         \sum_l  |\phi_l|^2  \delta_{jk}
         + \phi_j \phi_k^{\ast}
      \right\}
\nn\\&&
      + g_s
         \sum_{\alpha}\sum_{lm} \biggl[
            (F_{\alpha})_{jk} (F_{\alpha})_{lm} \phi_l^{\ast} \phi_m
\nn\\&&\quad
            +(F_{\alpha})_{jm} (F_{\alpha})_{lk} \left(
               \phi_l^{\ast} \phi_m
            \right)
         \biggr]
\nn\\&&
    - i \hbar \Omega \cdot \nabla \times {\bf r} \delta_{jk} ,
\label{eq:a}
\\
   B_{jk} &=&
      g_n \phi_j \phi_k
      +g_s
         \sum_{\alpha} \sum_{lm}
            (F_{\alpha})_{jl} \phi_l  (F_{\alpha})_{km} \phi_m.
\label{eq:b}
\end{eqnarray}
where $u_q({\bf r}, i)$ and $v_q({\bf r}, i)$ are
the $q$'th eigenfunctions with the spin component $i$ and
$\varepsilon_q$ corresponds to the $q$'th eigenvalue.
The normalization condition for the
$u_q({\bf r}, i)$ and $v_q({\bf r}, i)$ is
\begin{equation}
   \sum_i \int
   \left\{
       | u_q({\bf r}, i) |^2 - | v_q({\bf r}, i) |^2
   \right\} d{\bf r} = 1.
  \label{eq:normal}
\end{equation}

In our disc-shaped system,
the wavefunctions have the phase
$u_q({\bf r}, j) = u_q(r,j)e^{i \theta (  q_{\theta} + \beta_0 + j \beta)}$ and
$v_q({\bf r}, j) = v_q(r,j)e^{i \theta (  q_{\theta} - \beta_0 - j \beta)}$.
Then Eqs.\ (\ref{eq:bogo1} - \ref{eq:b}) are written as
\begin{eqnarray}
\lefteqn{
    \sum_j
        \{X_{ij}^{+} u_q(r,j) - 2 Z_{ij} v_q(r,j)\}
}
\nn\\
    &=&
    \left(
        \varepsilon_q + \hbar \Omega q_{\theta}
    \right) u_q(r,i)
\\
\lefteqn{
    \sum_j
    \{2 Z_{ij}^\ast u_q(r,j) - X_{ij}^{-} v_q(r,j)\}
}
\nn\\
    &=&
    \left(
        \varepsilon_q + \hbar \Omega q_\theta
    \right) v_q(r,i)
\label{eq:bogo2}
\end{eqnarray}
where
\begin{eqnarray}
    X_{ij}^{\pm} &=&
    [
        -C \left\{
            \frac{d^2}{dr^2} + \frac{1}{r}\frac{d}{dr}
            + \frac{(q_{\theta} \pm (\beta_0 + \beta j) )^2}{r^2}
        \right\}
\nn\\&&
        + V(r) - (\mu + j \mu^{\prime})
        + g_n \sum_k |\phi_k|^2
    ] \delta_{ij}
\nn\\&&
    + g_n ( \phi_j^{\ast} \phi_i ) \pm \hbar \Omega (\beta_0 + \beta j)
\delta_{ij}
\\
    Z_{ij} &=&
    g_n \phi_i \phi_j
    + g_s \sum_{\alpha}\sum_{kl} \phi_k \phi_l
    (F_{\alpha})_{il}
    (F_{\alpha})_{jk}.
\end{eqnarray}

The excitation energy $\varepsilon_q$ varies
as a function of the angular velocity $\Omega$
as determined
by Eqs.\ (\ref{eq:bogo1}) and (\ref{eq:bogo2}).
The critical values for local stability are defined by
\begin{eqnarray}
 \Omega_c^{+} &=&
     \min_{q_{\theta} > 0} \left( \varepsilon_q / q_\theta \right)
\label{omega_plus}
\\
 \Omega_c^{-} &=&
     \max_{q_{\theta} < 0} \left( \varepsilon_q / q_\theta \right).
\label{omega_minus}
\end{eqnarray}
The critical velocity $\Omega_c^{+}$ defined above corresponds to the
instability
of the surface excitations;
namely, the energy of the
excitation modes for ${q_{\theta} > 0}$
become negative (see the paper~\cite{isoshima2} by Isoshima and Machida for
the one-component BEC).
The critical velocity $\Omega_c^{-}$ corresponds to the local instability
where the core excitation modes with ${q_{\theta} < 0}$ become negative upon
varying $\Omega$. In the following, we determine $\Omega_c^{\pm}$,
yielding a stability region in the $\Omega$ vs $M/N$ plane where the all
excitation modes are positive definite for a given vortex configuration.
Namely, we evaluate
$\Omega_c^{\pm}$ for the vortex types (1,0,-1), (1,1,1), and (1,$x$,0)
as functions of $\Omega$ and $M/N$.

When there is a mode with $q_\theta = 0$ and $\varepsilon < 0$,
the system has an instability regardless of external rotation
and the definition of $\Omega_c^{\pm}$ becomes meaningless.
There are few modes found with $\varepsilon < 0$ for $q_\theta = 0$.
The value is at $-10^{-12} < \varepsilon < 0$.
There is also a complex mode with $q_\theta = 0$.
The real part is small (${\rm Re}(\varepsilon) < 10 \times 10^{-13}$)
while the imaginary part is of order $O(0.01)$.
We ignore them as the modes degenerate with the condensate.

\subsection{Local stability region}

Let us investigate $\Omega_c^{\pm}$ for the (1,0,-1), (1,$x$,0), and (1,1,1) vortices.
Figures \ref{fig:localomg}(a), \ref{fig:localomg}(b), and \ref{fig:localomg}(c)
show the local stability regions for each vortex where all the excitation
modes have positive eigenvalues.
Figure \ref{fig:shape_of_uv} shows
the radial shape and $q_\theta$ of the wavefunctions
which correspond to the critical value $\Omega_c^{\pm}$.

\subsubsection{Similarity to scalar system}

The $\Omega_c^{\pm}$ of the (1,0,-1) vortex for $M/N = \pm 1$ and
of the Alice vortex for $M/N = 1$ may be understood as those of a vortex in
a scalar BEC.
The $\Omega_c^{\pm}$ in Fig.\ \ref{fig:localomg}
reduces to those of the one-component
vortex system for $M/N=1$
because the fully polarized case is nothing but
a one-component BEC with a single vortex.
When we decrease $M/N$ towards $-1$, the (1,0,-1) vortex
becomes a single-component system again with the opposite winding number
and the critical $\Omega_c^{\pm}$'s have opposite signs.

As seen in Fig.\ \ref{fig:shape_of_uv},
the modes responsible for $\Omega_c^{+}$
are localized on the edge of the condensate ($r \simeq 3 \mu {\rm m}$)
and have a large angular momentum $q_{\theta}\simeq 7$
when $M/N > 0.9$ in both the (1,0,-1) and (1,$x$,0) vortices.
This implies a surface-instability mode
similar to the surface mode in a scalar vortex.
The mode remains at the edge of the condensate for $M/N > 0.8$.

The angular momenta of the modes of $\Omega_c^{-}$ are $-1$
over most of the regions: for  $0 < M/N \le 1$
in the (1,0,-1) vortex and for $-1 < M/N \le 1$ in the Alice vortex.
This mode is localized at the center when the condensate has a
large angular momentum
[$M/N  = 1$ in Figs.\ \ref{fig:shape_of_uv}(a),
\ref{fig:shape_of_uv}(b), \ref{fig:shape_of_uv}(c)].
This feature arises from the localized core state in a scalar vortex.

\subsubsection{Characteristic of spinor system}

\begin{figure}[htbp]
\begin{center}
\includegraphics[width=7cm]{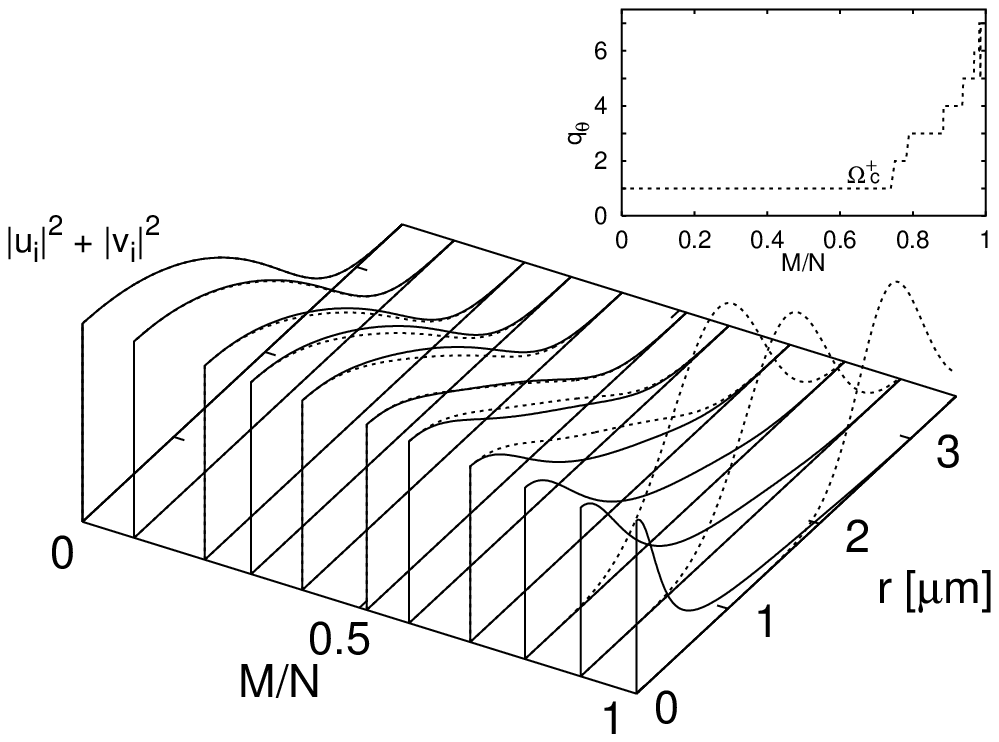}\\
\vspace{-0.5cm}
(a)

\includegraphics[width=7cm]{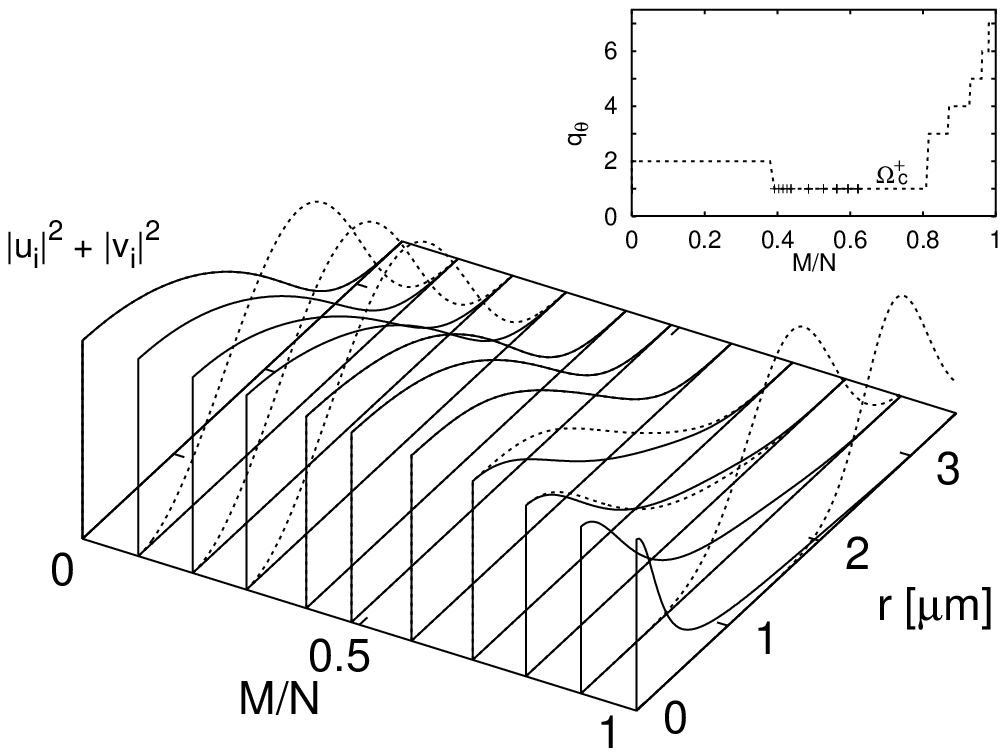}\\
\vspace{-0.5cm}
(b)

\includegraphics[width=7cm]{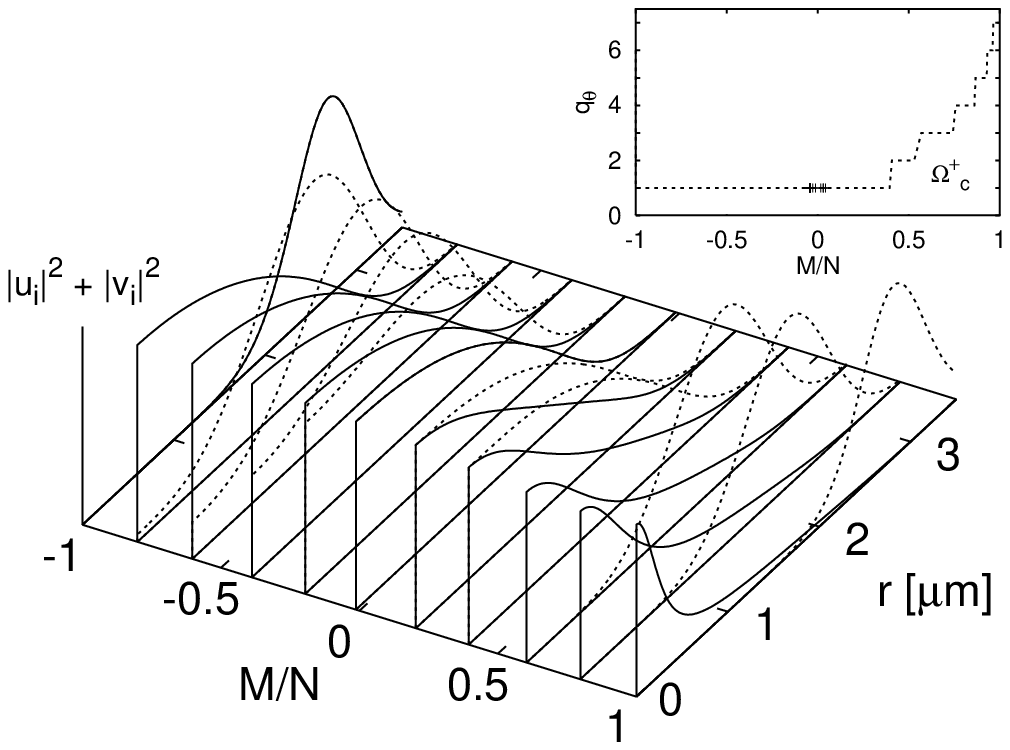}\\
\vspace{-0.5cm}
(c)

\end{center}
\caption{\label{fig:shape_of_uv}
Spatial variations of $\sum_i (|u_q(r,i)|^2 + |v_q(r,i)|^2)$
as a function of the radial direction  corresponding to the collective mode
at $\Omega_c^{\pm}$
for each $M/N$ is plotted.
The (dotted) lines corresponds to $\Omega_c^{-}$ ($\Omega_c^{+}$)
When the eigenfunctions $u_i$ and $v_i$ are complex number,
the sum of the real part
$\sum_i ({\rm Re}(u_q(r,i))^2 + {\rm Re}(v_q(r,i))^2)$
is plotted instead.
Inset of each figure shows the angular momentum index $q_{\theta}$ of
the mode corresponding to $\Omega_c^{+}$.
When $M/N = 1$, the mode of $\Omega_c^{+}$ has a peak at the edge of
the condensate ($r \sim 2 \mu {\rm m}$)
and the mode at $\Omega_c^{-}$ has a peak at the vortex center ($r=0$).
These reproduce the result of the scalar BEC.
When $M/N = 1$, $q_{\theta} = 7$ and it reduces to 1 as $M/N$ decreases.
(a) (1,0,-1) vortex in the ferromagnetic case.
(b)(1,0,-1) vortex in the antiferromagnetic case.
(c) (1,$x$,0) vortex in the nonmagnetic case.
}
\end{figure}

The spatial extent of the modes of $\Omega_c^{-}$,
which are localized around the
core for large $M/N$ expands as $M/N$ decreases.
This is shown for
$M/N  = 0$ in Figs.\ \ref{fig:shape_of_uv}(a)
and \ref{fig:shape_of_uv}(b), and
for $M/N  = -1$ in Figs.\ \ref{fig:shape_of_uv}(c).
This is common to both the (1,0,-1) and (1,$x$,0) vortices.
The wavefunction has a shape similar to that of the condensate
and a spin structure different from the condensate.
Thus this instability at $\Omega_c^{-}$ means a spin-flip instability.
This is also true for $\Omega_c^{+}$ of the (1,0,-1) vortex
for $-0.5 < M/N < 0.5$.
The shape of the wavefunctions is similar to that of the condensate.
This is not the situation in the antiferromagnetic case.
The wavefunction of $\Omega^{+}_c$ indicates a single vortex of the -1
component
rather than the non-vortex -1 component,
as shown in Fig.\ \ref{fig:shape_of_uv} (b).

According to Figs.\ \ref{fig:localomg} (a-c),
the critical values $\Omega_c^{+}$ and $\Omega_c^{-}$ of the (1,0,-1) vortex
are close to each other for $-0.5 < M/N < 0.5$.
There is no range of $\Omega$ to stabilize the system in the ferromagnetic case
because $\Omega^{-}_c > \Omega^{+}_c$.
This range is quite narrow in the antiferromagnetic case
({\it e.g.}, $0.261 < |\Omega| < 0.263$ for $M/N = \pm 0.2$).

The (1,$x$,0) vortex reduces to the vortexfree system near $M/N = -1$.
As shown in Figs.\ \ref{fig:localomg}(a) and \ref{fig:localomg}(b),
there is an instability for $\Omega = 0$ unlike the vortexfree scalar system.
Figure \ref{fig:shape_of_uv}(c) shows the shape of the wavefunctions $u$ and $v$
of $\Omega_c^{-}$.
The shape resembles that of the condensate.
This instability indicates that the whole condensate may have the +1 component
without winding number.
The winding-number combination of the  (1,$x$,0) vortex turns into
(0,0,0).
When $\mu^{\prime}$ is large and negative,
$\Omega^{+}_{c}$ and  $\Omega^{-}_{c}$ become about $0.5$ and $-0.3$
respectively.
Therefore, the system is stable for $\Omega = 0$ just like the
vortexfree scalar system.
The $\Omega^{\pm}_{c}$ shift, depending on $\mu^{\prime}$, even when $M/N$
has the fixed value $-1$.
This is because our calculation
does not include the magnetization from the non-condensate.

\subsubsection{Complex eigenvalues}

Complex eigenvalues emerge when two levels with the opposite
angular momenta and the same eigenvalues happen to appear.
The complex eigenvalues of Bogoliubov equations have been reported for
multiply quantized vortex of scalar BEC~\cite{pu}
and vortex states of binary BEC~\cite{skryabin}.
When the eigenvalue $\varepsilon_q$ has a complex value,
the left-hand side of Eq.~(\ref{eq:normal}) becomes zero.
We still adopt the real part of such eigenvalues because
continuity of 
the real eigenvalues and the real part of complex eigenvalues
as shown in Fig.\ \ref{fig:localomg}
seems physically meaningful.


\subsubsection{The (1,1,1) vortex}

As for the (1,1,1) vortex, neither the ferromagnetic nor
the nonmagnetic states can assume
intermediate values of magnetization and $M/N$ becomes $\pm 1$.
Even at each magnetization, $\Omega_c^{\pm}$ varies with the
chemical potential $\mu^{\prime}$.
The maximum value of $\Omega_c^{-}$ is about 1.8 (in trap units).
The wavefunction corresponding to $\Omega_c^{-}$ is localized at the center
and has a spin component different from that of the condensate.
The minimum value of $\Omega^{+}_c$ is about 0.4.
Because $\Omega_c^{-}$ is always larger, the (1,1,1) vortex is always
unstable in the sense of local stability.

This is also the case for the antiferromagnetic system.
$\Omega^{-}_c \simeq 1.8$ and $\Omega^{+}_c \simeq 0.4$ mean that the (1,1,1)
vortex is locally unstable.
When the chemical potential $\mu^{\prime}$ is large enough,
these critical values $\Omega^{\pm}_c$ reduce to
 $\Omega^{+}_{c} = 0.6$ and $\Omega^{-}_{c} = 0.3$.

%
%
%

\section{Discussion}

We have determined the vortex phase diagram
in the plane of external rotation frequency $\Omega$ vs
relative magnetization $M/N$
of a spinor BEC system
for the cases of antiferromagnetic ($g_s>0$), nonmagnetic ($g_s=0$),
and ferromagnetic ($g_s<0$) interactions.
By solving the extended Gross-Pitaevskii equation
for the  spinor F=1 BEC with the
three components ($m_F=1,0,-1$), we have
investigated the relative stability of the possible
axisymmetric vortices
which are classified according to their winding numbers for each component;
namely, (1,0,-1), (1,$x$,0), and
(1,1,1) for ($\phi_1, \phi_0,\phi_{-1}$).
The excitation spectra are also studied to find
the stability regions for each vortex in the  $\Omega$ and $M/N$ plane.
This allows us to estimate the vortex nucleation frequency in the actual
experiment (for details, see Refs. \cite{isoshima2,mizushima} for
the correspondence between local stability and nucleation frequency).

Yip~\cite{yip} has studied the two non-axisymmetric vortices in addition
to the present axisymmetric vortices within the GP equation
under a particular rotation frequency ($\Omega\sim 0.4$ in our notation)
for the antiferromagnetic case only. This corresponds to
Fig. \ref{fig:global}(a) for $\Omega\sim 0.4$.
Although the lower $M/N$ region ($0<M/N<0.2$) is occupied by a
non-axisymmetric
vortex, in the  remaining region the (1,$x$,0) vortex is stabilized
over the non-axisymmetric vortex,
coinciding with our result.
We believe that the present phase diagram remains valid
over a broad region
even taking into account non-axisymmetric vortices.
This is because a large part of Yip's
phase diagram is covered by axisymmetric vortices (see Fig.1 in Yip's
paper \cite{yip}).
In the future we aim to 
take into account the non-axisymmetric vortex in
addition to the present axisymmetric ones.
This kind of a calculation must be
performed with the help of the
excitation spectrum, which signals the instability towards a more stable
vortex configuration,
for each vortex and is needed to predict the experimentally realized
vortices in a spinor BEC.

\section*{Acknowledgements}
The authors thank T. Ohmi, T. Mizushima, T. Kita,
M.~M.~Salomaa, 
S.~M.~M.~Virtanen, and T.~Simula
for useful discussions.




\begin{thebibliography}{1}


\bibitem{firstRb}
M.~H. Anderson, J.~R. Ensher, M.~R. Matthews, C.~E. Wieman and E. Cornell,
  Science {\bf 269}, 198 (1995).

\bibitem{hulet}
C.~C. Bradley, C.~A. Sackett, J.~J. Tollett and R.~G. Hulet,  Phys. Rev. Lett.
  {\bf 75}, 1687 (1995).

\bibitem{ketterle}
K.~B. Davis, M.-O. Mewes, M.~R. Andrews, N.~J. van Druten, D.~S. Durfee, D.~M.
Kurn and W. Ketterle,  Phys. Rev. Lett. {\bf 75}, 3969 (1995).

\bibitem{pethick}
C.~J.~Pethick and H.~Smith,
{\em Bose-Einstein Condensation in Dilute Gases}
(Cambridge University Press, 2002).

\bibitem{cornish}
 S. L. Cornish, N. R. Claussen, J. L. Roberts, E. A. Cornell and C. E.
 Wieman,  Phys. Rev. Lett. {\bf 85}, 1795 (2000).

\bibitem{fried}
D. G. Fried, T. C. Killian, L. Willmann, D. Landhuis, S. C. Moss, D. Kleppner
and T. J. Greytak,   Phys. Rev. Lett. {\bf 81}, 3811(1998).

 \bibitem{robert}
A. Robert, O. Sirjean, A. Browaeys, J. Poupard, S. Nowak, D. Boiron, C. I.
Westbrook  and
A. Aspect,  Science {\bf 292}, 461 (2001).

\bibitem{modugno}
G. Modugno, G. Ferrari, G. Roati, R.J. Breecha, A. Simoni and
M. Inguscio,  Science {\bf 294}, 1320 (2001).


\bibitem{stenger}
J. Stenger, S. Inouye, D.~M. Stamper-Kurn, H.-J. Miesner, A.~P.
Chikkatur and W. Ketterle, Nature {\bf 369},  345 (1998).
H.-J. Miesner, D.~M. Stamper-Kurn, J. Stenger, S. Inouye,
A.~P. Chikkatur and W. Ketterle,  Phys. Rev. Lett. {\bf 82},  2228(1999).
D.~M. Stamper-Kurn, H.-J. Miesner, A.~P. Chikkatur, S. Inouye,
J. Stenger and W. Ketterle,  Phys. Rev. Lett. {\bf 83},  661 (1999).


\bibitem{barrett}
M. Barrett, J. Sauer and M.S. Chapman, Phys. Rev. Lett. {\bf 87}, 010404
(2001).

\bibitem{ohmi}
T. Ohmi and K. Machida, J. Phys. Soc. Jpn. {\bf 67} (1998) 1822.

\bibitem{ho}
T.-L. Ho, Phys. Rev. Lett. {\bf 81},  742. (1998).


\bibitem{salomaa_vtx}
M.~M.~Salomaa and G.~E.~Volovik, Rev. Mod. Phys. {\bf 59}, 533 (1987).

\bibitem{salomaa_wall_b}
M.~M.~Salomaa and G.~E.~Volovik, Phys. Rev. B. {\bf 37}, 9298 (1988).

\bibitem{salomaa_wall_a}
M.~M.~Salomaa and G.~E.~Volovik, J. Low Temp. Phys. {\bf 74}, 319 (1989).




\bibitem{yip}
S.-K. Yip, Phys. Rev. Lett. {\bf 83},  4677 (1999).


\bibitem{isoshima}
T. Isoshima, K. Machida and T. Ohmi, J. Phys. Soc. Jpn. {\bf 70}, 1604 (2001).




\bibitem{leonhardt}
U. Leonhardt and G. E. Volovik, JETP Lett. {\bf 72},  46 (2000).



\bibitem{stoof}
H.T.C. Stoof, cond-mat/0002375.


\bibitem{marzlin}
Karl-Peter Marzlin, Weiping Zhang and Barry C. Sanders,
Phys. Rev. A {\bf 62}, 13602 (2000).


\bibitem{zhou}
Fei Zhou, Phys. Rev. Lett. {\bf 87}, 080401 (2001).

\bibitem{martikainen}
J.-P. Martikainen and K.-A. Suominen,  cond-mat/0106013.


\bibitem{spindomain}
T. Isoshima, K. Machida and T. Ohmi, Phys. Rev. A {\bf 60},  4857 (1999).


\bibitem{isoshimanakahara1}
T. Isoshima, M. Nakahara, T. Ohmi and K. Machida,
Phys. Rev. A {\bf 61},  063610 (2000).


\bibitem{isoshimanakahara2}
M. Nakahara, T. Isoshima, K. Machida, S. Ogawa and T. Ohmi,
Physica B {\bf 284-288},  17 (2000).


\bibitem{double}
T. Isoshima, T. Ohmi and K. Machida,
J. Phys. Soc. Jpn. {\bf 69}, 3864 (2000).


\bibitem{isoshima1}
T. Isoshima and K. Machida, J. Phys. Soc. Jpn. {\bf 66},  3502 (1997).

 \bibitem{isoshima2}
T. Isoshima and K. Machida, J. Phys. Soc. Jpn. {\bf 68},  487 (1999).

 \bibitem{isoshima_T}
T. Isoshima and K. Machida, Phys. Rev. A {\bf 59},  2203 (1999).

 \bibitem{isoshima4}
T. Isoshima and K. Machida, Phys. Rev. A {\bf 60},  3313 (1999).

 \bibitem{mizushima}
T. Mizushima, T. Isoshima and K. Machida, Phys. Rev. A {\bf 64},  043610
(2001).

 \bibitem{virtanen}
S. M. M. Virtanen, T. P. Simula and M. M. Salomaa,
Phys. Rev. Lett. {\bf 86}, 2704 (2001).

\bibitem{pu}
H. Pu, C.~K.~Law, J.~H.~Eberly, and N.~P.~Bigelow,
Phys. Rev. A {\bf 59}, 1533 (1999).

\bibitem{skryabin}
D. V. Skryabin, Phys. Rev. A {\bf 63}, 013602 (2000).




\end{thebibliography}
\end{document}